\documentclass[twocolumn,showpacs,prl,superscriptaddress]{revtex4-1}
\usepackage{amsfonts}
\usepackage{amsmath}
\usepackage{amssymb}
\usepackage{graphicx}
\usepackage{natbib}
\usepackage[colorlinks=true,citecolor=blue,urlcolor=blue]{hyperref}
\usepackage{color}

\begin{document}

\title{Long-time dynamics of quantum chains: transfer-matrix renormalization
group and entanglement of the maximal eigenvector}
\date{\today }
\author{Yu-Kun Huang}
\email{ykln@mail.nju.edu.tw}
\affiliation{Graduate School of Engineering Science and Technology,\\
Nan Jeon University of Science and Technology, Tainan 73746, Taiwan}
\author{Pochung Chen}
\email{pcchen@phys.nthu.edu.tw}
\affiliation{Department of Physics and Frontier Research Center on Fundamental and
Applied Sciences of Matters, \\
National Tsing Hua University, Hsinchu 30013, Taiwan}
\author{Ying-Jer Kao}
\email{yjkao@phys.ntu.edu.tw}
\affiliation{Department of Physics and Center for Advanced Study of Theoretical Science,\\
National Taiwan University, No. 1, Sec. 4, Roosevelt Rd., Taipei 10607,
Taiwan}
\author{Tao Xiang }
\email{txiang@iphy.ac.cn}
\affiliation{Institute of Physics, Chinese Academy of Sciences, P.O. Box 603, Beijing
100190, China}

\begin{abstract}
By using a different quantum-to-classical mapping from the Trotter-Suzuki
decomposition, we identify the entanglement structure of the maximal
eigenvectors for the associated quantum transfer matrix. This observation
provides a deeper insight into the problem of linear growth of the
entanglement entropy in time evolution using conventional methods. Based on
this observation, we propose a general method for arbitrary temperatures
using the biorthonormal transfer-matrix renormalization group. Our method
exhibits a competitive accuracy with a much cheaper computational cost in
comparison with two recent proposed methods for long-time dynamics based on
a folding algorithm [Phys. Rev. Lett. \textbf{102}, 240603 (2009)] and a
modified time-dependent density-matrix renormalization group [Phys. Rev.
Lett. \textbf{108}, 227206 (2012)].
\end{abstract}

\maketitle


The response function of a system subject to an external perturbation is a
fundamental quantity to understand the mechanism inside a strongly
interacting system. In particular, the time-dependent correlation function,
whose Fourier transformation gives the spectral information about the
system, can be measured in experiments. Therefore, it is important to be
able to study the time-dependent correlation function with high accuracy. 
For one-dimensional (1D) systems, the density matrix renormalization group
(DMRG) is an efficient algorithm to accurately obtain the ground state of
interacting quantum models\cite{SWhite1992}, and has been extended to
address real-time dynamics \cite{Vidal1, *SWhite2004td,*Vidal2} and the
computation of thermodynamic quantities\cite%
{AFeiguin2005,*Barthel2009,*TNishino1995,*Bursill1996vn,*XWang1997,*Shibata1}%
. Long-time dynamics, however, remains difficult due to the linear growth of
the entanglement entropy as the state evolves in time\cite%
{Barthel2009,Sirker:2005tg}. Recently, two schemes have been proposed which
allow numerical stable computation of the long-time dynamics. The folding
scheme uses a more efficient representation of the entanglement structure in
the tensor network obtained from the Trotter-Suzuki decomposition, by
folding the network in the time direction prior to contraction\cite%
{Banuls1,*Banuls2}. On the other hand, a modified finite-temperature
time-dependent DMRG (tDMRG) scheme exploits the freedom of applying unitary
transformations in the ancilla space \cite{Karrasch2012,*Barthel2012}. 
Although these methods can reach longer time scale than before, it is not
clear the reason why they should work and whether if they can be extended to
generic quantum models.

In this paper, we construct a single-site quantum transfer matrix
(QTM) using an alternative quantum-to-classical mapping\cite{Sirker:2002kl}
for generic quantum models. Analyzing this QTM gives us a clear picture of
the entanglement structure for the QTM. This not only provides a deeper
insight into the problem of entanglement growth during the time evolution,
but also a heuristic argument on why the conventional schemes break down at
long time scale and why the folding and the modified tDMRG schemes work.
Furthermore, we propose an alternative scheme using the biorthonormal
transfer-matrix DMRG (BTMRG) \cite{YHuang2011a,*YHuang2011b,*YHuang2012} by
partitioning the QTM into system and environment blocks according to the
entanglement structure. We show that our scheme exhibits the same level of
accuracy with a much cheaper numerical cost.

\begin{figure}[tb]
\includegraphics[width=7cm,clip]{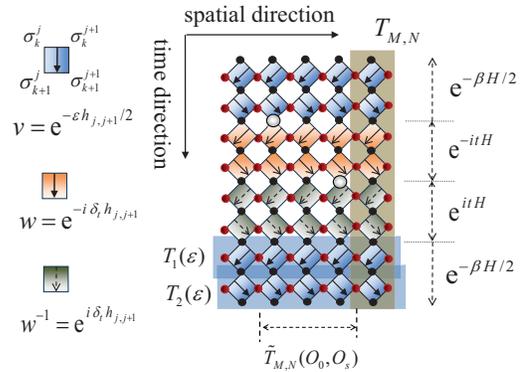}\vspace{-0.3cm} 
\caption{(Color online) Graphical representation of the evolution matrices $T_{1,2} (\varepsilon )$ and the quantum transfer matrix $T_{M,N}$.
$\sigma_k^j$ is a basis state at site $j$ and time (or imaginary time) $k$.
$\widetilde{T}_{M,N}(O_{0},O_{s})$ is a product of $s+1$ transfer matrices 
where the two empty circles represents two operators at different site and 
different time. Detailed definitions of these matrices are given in 
Supplementary Material S1.
} \label{fig1}
\end{figure}

We start from a 1D quantum system with $L$ sites and a Hamiltonian $%
H=\sum_{j=1}^{L}{h}_{j,j+1}$ with nearest-neighbor interactions and periodic
boundary conditions. The time dependent correlation function for an operator 
$O$ between site $s$ and site $0$ at finite temperature can be expressed as 
\begin{equation}
\left\langle O_{s}(t)O_{0}(0)\right\rangle =\frac{1}{Z}\text{Tr}\left( \text{%
e}^{-\beta H}\text{e}^{itH}O_{s}\text{e}^{-itH}O_{0}\right) ,
\label{eq:AutoCF}
\end{equation}%
where $Z=\text{Tr(}$e$^{-\beta H})$ is the partition function and $\beta
=1/T $ is the inverse temperature.

We employ a quantum-to-classical mapping that decomposes the operator e$%
^{-\beta {H}}$ into 
\begin{equation}
\text{e}^{-\beta H}\approx \lim_{M\rightarrow \infty }\left[ \emph{T}%
_{1}(\varepsilon )\emph{T}_{2}(\varepsilon )\right] ^{M/2},
\end{equation}%
where $T_{1,2}(\varepsilon )=T_{R,L}$e$^{-\varepsilon H}$ is a row-to-row
evolution operator, and $T_{R,L}$ are the right and left shift operators
which are defined by the product of a 45$^{\circ }$ clockwise and
counter-clockwise rotated local operator $\nu \equiv $ e$^{-\varepsilon
h_{j,j+1}}$, respectively. $\varepsilon =\beta /M$ is the imaginary-time
step and $M$ is the Trotter number in the imaginary-time direction \cite%
{Sirker:2002kl}. A similar decomposition can be applied to the operators e$%
^{-itH}$ and e$^{itH}$ by replacing $\nu $ with the complex local operators $%
w\equiv $ e$^{-i\delta _{t}h_{j,j+1}}$ and $w^{-1}\equiv $ e$^{i\delta
_{t}h_{j,j+1}}$, respectively. Here $\delta _{t}=t/N$ and $N$ is the Trotter
number in the time direction. From these decompositions a column-to-column
QTM, $T_{M,N}$, can be defined. A graphical representation of $T_{M,N}$ is
shown in Fig. \ref{fig1} (a more detailed definition of $T_{M,N}$ is given
in Supplementary Material S1).

In the thermodynamic limit, the correlation functions in Eq. (\ref{eq:AutoCF}%
) can be determined by the maximal eigenvalue $\Lambda _{0}$\ and the
corresponding left $\langle \psi _{M,N}^{l}|$ and right $|\psi
_{M,N}^{r}\rangle $ eigenvectors of $T_{M,N}$: 
\begin{equation}
\left\langle O_{s}(t)O_{0}(0)\right\rangle =\frac{\langle \psi _{M,N}^{l}|%
\widetilde{T}_{M,N}(O_{0},O_{s})\left\vert \psi _{M,N}^{r}\right\rangle }{%
\Lambda _{0}^{s+1}}.  \label{eq:CFtmrg}
\end{equation}%
Here $\langle \psi _{M,N}^{l}|\psi _{M,N}^{r}\rangle =1$ is implied and $%
\widetilde{T}_{M,N}(O_{0},O_{s})\equiv T_{M,N}(O_{0})T_{M,N}^{\text{ }%
s-1}T_{M,N}(O_{s})$ is a product of $s+1$ transfer matrices, where $%
T_{M,N}(O_{j})$ is a modified transfer matrix containing operator $O_{j}$ at
site $j=0$ and $s$ (see Fig. \ref{fig1}).


\begin{figure}[tb]
\includegraphics[width=8cm,clip]{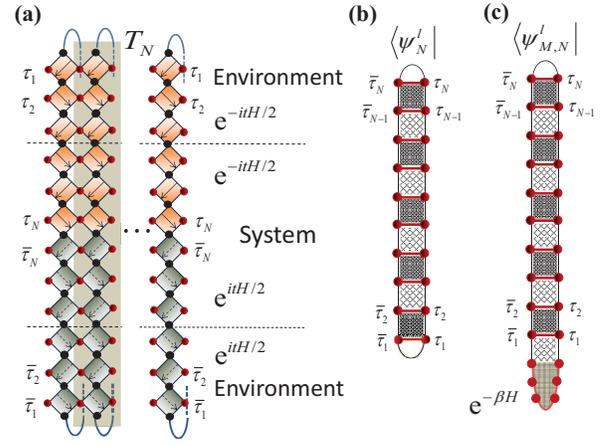} \vspace{0cm} 
\caption{(Color online)
(a) Graphical representation of the transfer matrix $T_N$. $\tau_n$ and $\overline{\tau}_n$ represent the time along the forward and backward evolution directions, respectively.
(b) Schematic plot of the entanglement structure for the maximal left eigenvector of $T_{N}$. $\langle \psi_{N}^{l}|$ is an entangled state of singlet pairs between $\tau _{n}$ and $\overline{\tau }_{n}$. The two parts separated by a dark shadow has stronger entanglement than that separated by a light shadow. The right maximal eigenvector $|\psi_{N}^{r}\rangle $ has a similar entanglement structure with the dark and light shadows interchanged.
The new TMRG scheme bi-partitions the QTM into e$^{itH/2}$e$^{-itH/2}$\ and e$^{-itH/2}$e$^{itH/2}$ such that the maximally entangled pair is locked within each block.
(c) The maximal eigenvectors of $T_{M,N}$ at finite temperature have similar structures.
} \label{fig2}
\end{figure}


In order to evaluate the long time correlation function accurately, one
needs to understand how the entanglement is built up in the maximal
eigenvectors of $T_{M,N}$ during the time evolution. It is instructive to
first consider the infinite-temperature case where the thermal density
matrix e$^{-\beta H}$ becomes an identity and $T_{M,N}$ associated with its
dual eigenvectors $\langle \psi _{M,N}^{l}|$ and $|\psi _{M,N}^{r}\rangle $
are reduced to an $M$-independent matrix $T_{N}$ associated with $\langle
\psi _{N}^{l}|$ and $|\psi _{N}^{r}\rangle $ (Fig. \ref{fig2}(a)). For the
convenience in the discussion below, we use $\tau _{n}^{\ast }$ to denote a
pair of virtual states $(\tau _{n},\overline{\tau }_{n})$, $n=1,...N,$ and $%
|\tau _{n}^{\ast }\rangle =\sum_{\tau _{n}\overline{\tau }_{n}}\delta _{\tau
_{n},\overline{\tau }_{n}}|\tau _{n},\overline{\tau }_{n}\rangle $ to denote
a maximally entangled state of $\tau _{n}$ and $\overline{\tau }_{n}$. An
important property of $T_{n}$, as shown in the supplementary material, is
that the contraction of its left maximal eigenvector $\langle \psi _{n}^{l}|$
with $|\tau _{n}^{\ast }\rangle $ satisfies the following equation 
\begin{equation}
\langle \psi _{n}^{l}|\tau _{n}^{\ast }\rangle =\langle \psi
_{n-1}^{l}|=\langle \psi _{n-2}^{l}|\langle \tau _{n-1}^{\ast }|
\label{eq:C}
\end{equation}%
for an even $n$. A similar equation holds for the right eigenvector $|\psi
_{n}^{r}\rangle $ but with odd $n$. Eq.~\ref{eq:C} means that the
contraction of the left eigenvector $\langle \psi _{n}^{l}|$ with a
maximally entangled pair $|\tau _{n}^{\ast }\rangle $ produces another
separated maximally entangled pair $|\tau _{n-1}^{\ast }\rangle $ if $n$ is
even. This implies that the virtual basis states $\tau _{n}$ and $\overline{%
\tau }_{n}$ at the same forward and backward real time $n$ have strong
tendency to form a singlet pair. Thus the left maximal eigenvector should
have an entanglement structure as depicted in Fig. \ref{fig2}(b), where the
red bonds denote singlet pair states, and the upper and lower parts
separated by a dark (light) shadow have stronger (weaker) entanglement. A
similar entanglement structure exists for the right maximal eigenvector $%
|\psi _{n}^{r}\rangle $, but the dark and light shadows are interchanged.

As taking into account the finite temperature effect, the QTM $T_{M,N}$ is
obtained by simply adding two blocks of thermal operator at the top and
bottom of $T_{N}$ as shown in Fig. \ref{fig1}. Following the derivation of
Eq. (\ref{eq:C}) (see Supplementary Material S2), we can obtain%
\begin{equation}
\langle \psi _{M,n}^{l}|\tau _{n}^{\ast }\rangle =\langle \psi
_{M,n-1}^{l}|=\langle \psi _{M,n-2}^{l}|\langle \tau _{n-1}^{\ast }|
\end{equation}%
for an even real-time $n$, and the eigenvector $|\psi _{M,n}^{r}\rangle $
has a similar property for odd $n$. Here $\langle \psi _{M,n}^{l}|$ and $%
|\psi _{M,n}^{r}\rangle $ denote the left and right maximal eigenvectors of $%
T_{M,n}$ with $M$ fixed and $n=1,...,N$. Thus, the eigenvectors of $T_{M,N}$
should have a slower increasing entanglement between the block involving
imaginary-time and the block involving real-time. Furthermore, the
eigenvectors of $T_{M,N}$ should have a similar entanglement structure for
the block involving real-time as the structure of the eigenvectors of $T_{N}$
(see Fig. \ref{fig2}(c)).

\begin{figure}[tb]
\includegraphics[width=8cm,clip]{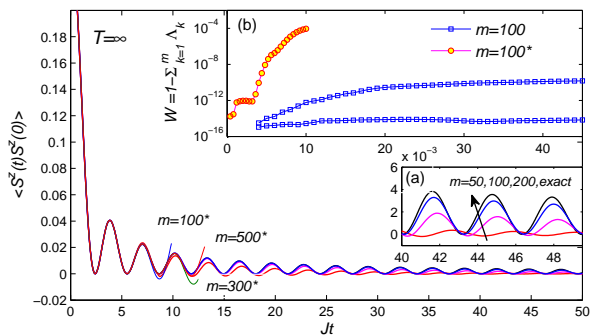} \vspace{-0.2cm} 
\caption{(Color online)
Time-dependent autocorrelation $\langle S^{z}(t)S^{z}(0)\rangle$ for the Heisenberg model (\ref{eq:Haml}) with $\Delta =0$ at infinite temperature.
$m$ is the number of basis states retained in the calculation. The results
The data with $m$ marked by asterisk are obtained by the conventional TMRG scheme.
Inset (a): Blow-up of the autocorrelation in the long time region.
Inset (b): Comparison of the discarded weight $W=1-\sum_{k=1}^{m}\Lambda _{k}$ obtained by the BTMRG with that obtained by the conventional TMRG.
The two $W$-curves for our scheme correspond to the cutting boundary through the dark shadow with strong entanglement and through the light shadow with weak entanglement, respectively.
} \label{fig3}
\end{figure}

The above discussion suggests that in order to suppress the growth of the
entanglement with time evolution, we should redefine $T_{N}$ on a folded
lattice where $\tau _{n}$ and $\overline{\tau }_{n}$ at real time $n$ are
merged into a single site, as shown in Fig. \ref{fig2}(b). For $T_{M,N}$ on
a folded lattice, the additional sites $\tau _{m}$ and $\overline{\tau }_{m}$
for imaginary time are also merged, but the folded block e$^{-\beta {H}}$
is regarded as a heat bath\textbf{\ }(Fig. \ref{fig2}(c)). Thus the strong
entanglement between $\tau _{N}$ and $\overline{\tau }_{N}$ is confined
within each time step and will not proliferate with time evolution. This can
avoid the linear growth problem of entanglement with time in the
conventional TMRG schemes\cite{Sirker:2005tg}.

Based on the above argument, we propose to do the BTMRG calculation by
collecting a segment of local matrices around the center of the virtual
lattice as the system block and the rest of local tensors as the environment
block (Fig. \ref{fig2}(a)). In the infinite temperature limit, this is
equivalent to dividing the quantum transfer matrix into $e^{itH/2}e^{-itH/2}$
and $e^{-itH/2}e^{itH/2}$ parts. At each iteration, both system and
environment blocks are enlarged by adding two different rotated matrices $%
w^{,}s$ ($w^{-1,}s$) at the upper (lower) boundary site between the two
blocks. For the finite-temperature calculation, we first cool down the
temperature to a desired value by taking the imaginary time evolution, and
then take the real time evolution by embedding $T_{M}$ into the environment
block. in this case, the system and environment blocks are equivalent to $%
e^{itH/2}e^{-itH/2}$ and $e^{-itH/2}e^{-\beta H}e^{itH/2}$, respectively.

To test the method, we calculate the longitudinal spin autocorrelation $%
\langle S^{z}(t)S^{z}(0)\rangle $ for the anisotropic spin-1/2 Heisenberg
model defined by the Hamiltonian 
\begin{equation}
H=\sum_{j}J(S_{j}^{x}S_{j+1}^{x}+S_{j}^{y}S_{j+1}^{y}+\Delta
S_{j}^{z}S_{j+1}^{z}).  \label{eq:Haml}
\end{equation}%
When $\Delta =0$, this model is equivalent to the spinless free fermion
model where an exact result for the autocorrelation is available \cite%
{Niemeijer1967}. We use the BTMRG to evaluate the maximal eigenvectors of $%
T_{M,N}$ \cite{YHuang2011a,*YHuang2011b}. By properly choosing the dual
biorthonormal bases, the BTMRG can achieve significant improvement of
numerical stability over the conventional TMRG.

Fig.~\ref{fig3} compares the results of the autocorrelation for the above
above model up to $Jt=50$ with $\delta_{t}=0.05$ and $\Delta =0$ at infinite
temperature obtained from both the BTMRG and the conventional TMRG. We find
that the time scale that can be reached by the conventional TMRG with
reliable accuracy is quite small, due to the fast increase of entanglement
entropy with evolving time. However, our BTMRG calculation can reach much
longer time scale with high precision. In particular, as shown in Fig.~\ref%
{fig3}(b) the truncation error $W=1-\sum_{k=1}^{m}\Lambda _{k}$ increases
very rapidly with increasing time scale in the TMRG calculation. In
contrast, in our BTMRG calculation $W$ saturates asymptotically. This shows
that the entanglement builds up mainly between the operators e$^{-itH}$\ and
e$^{itH}$. We also calculate the real-time dynamics for several spin-1/2
chains at various temperatures. The results are consistent with our
prediction for the entanglement structure at finite temperature and the
dynamics can be reliably evaluated on a time scale far longer than the
previous TMRG calculations (see the supplementary material).

\begin{figure}[tb]
\includegraphics[width=8cm,clip]{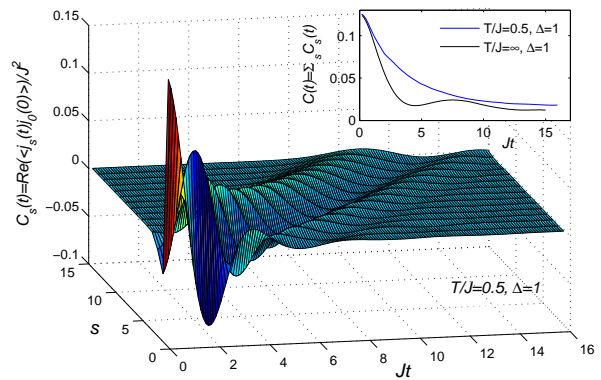} \vspace{-0.3cm} 
\caption{(Color online)
The correlation function $C_{s}(t)$ for the Heisenberg model with $\Delta =1$ and $s=0-15$ at $T/J=0.5$.
Inset: Current correlation function $C(t)$ for $\Delta =1$ at $T/J=0.5$ and $T=\infty$, respectively.
} \label{fig4}
\end{figure}

An advantage of our BTMRG scheme is its competitive accuracy with much
cheaper computational cost. The computational cost of the BTMRG scales with
the basis states retained as $O(m^{3})$, one order less than in the folding
algorithm as well as the tDMRG using matrix product states where the
computational cost scales as $O(m^{4})$. Moreover, our method works directly
in the thermodynamic limit and there is no finite lattice size effect. The
most challenging problem in our method is the calculation of long distance
correlation functions, since it involves a calculation of the product of
many transfer matrices $\widetilde{T}_{M,N}(O_{0},O_{s})$ (see Fig. \ref%
{fig1}).

In our algorithm, this wide transfer matrix $\widetilde{T}%
_{M,N}(O_{0},O_{s}) $ is approximated by a matrix-product operator (MPO) and
is updated in a similar way as the evolution of the matrix product state in
the modified tDMRG\cite{Karrasch2012,*Barthel2012} (see the supplementary
material). Our method, however, differs from the modified tDMRG in three
aspects: \textbf{(}i\textbf{)} The infinite-length MPO used in tDMRG for the
correlator is replaced by a $(s+1)$-length MPO representing the transfer
matrix $\widetilde{T}_{M,N}(O_{0},O_{s})$. This leads to dramatic reduction
of computational storage and time. \textbf{(}ii\textbf{)} The left (right)
bond of the first (last) MPO matrix is obtained by projecting the matrix
from the left (right) onto the reduced basis states describing the left
(right) maximal eigenvector. Since the number of state kept, $m$, is usually
small due to our special bi-partitioning scheme, the bond dimension $\chi $
of the MPO matrices is typically smaller than that in tDMRG. \textbf{(}iii%
\textbf{)} The correlator is obtained by contracting the $(s+1)$-length MPO
with the left and right maximal eigenvectors at the left and right bonds
respectively. These advantages can be seen from the calculation of the
current-current correlation function\cite%
{Karrasch2012,Sirker:2009hc,*Sirker:2011ST,*Jesenko2011}%
\begin{equation}
C_{s}(t)=\langle j_{s}(t)j_{0}(0)\rangle ,
\end{equation}%
where $j_{s}=-\frac{iJ}{2}(S_{s}^{+}S_{s+1}^{-}-S_{s+1}^{+}S_{s}^{-})$. The
long time limit of $C(t)=\sum_{s}C_{s}(t)$ gives the value of Drude weight.
Fig.~\ref{fig4} shows $C_{s}(t)$ for the Heisenberg model (\ref{eq:Haml})
with $\Delta =1$ and $s=0-15$ at $T/J=0.5$. The result is obtained by fixing
the truncation error less than $W=1\times 10^{-11}$ for the maximal
eigenvectors (the number of basis states is typically around $m=250$) and
less than $W=1\times 10^{-6}$ for the evolution of matrix product operators
with the bond dimension around $\chi =500$. The figure clearly shows that
the large distance and long-time dynamics can be accurately determined.
Another notable advantage of our method is that in the determination of
these correlation functions, the left and right maximal eigenvectors need to
be evaluated just once. The inset of Fig. \ref{fig4} shows the spatial
integrated current correlation function $C(t)$ up to time scale where $C(t)$
saturates for $\Delta=1$ at $T/J=0.5$ and $T=\infty $, which agree well with
the results given in Refs.~\onlinecite{Karrasch2012,Sirker:2009hc}.

To summarize, we identify the entanglement structure of the maximal
eigenvectors of the QTM proposed in \cite{Sirker:2002kl}. It reveals the
origin of the entanglement growth during the time evolution and the reason
why the recent algorithms work \cite{Banuls1,Karrasch2012}. On the basis of
this picture, we propose an alternative method based on the BTMRG approach
by bi-partitioning the system and the environment blocks according to the
entanglement structure. Our approach provides an very efficient tool for
studying finite-temperature dynamics of 1D quantum lattice models.

\begin{acknowledgments}
Y.-K.H. acknowledges the support by NSC in Taiwan through Grants No.
101-2112-M-232-001.
\end{acknowledgments}

\bibliographystyle{apsrev4-1}
\bibliography{BTMRG}

\end{document}


\title{Supplementary Material for \textquotedblleft Long-time dynamics of
quantum chains: transfer-matrix renormalization group and entanglement of
the maximal eigenvector\textquotedblright}
\date{\today }
\author{Yu-Kun Huang}
\affiliation{Graduate School of Engineering Science and Technology,\\
Nan Jeon University of Science and Technology, Tainan 73746, Taiwan}
\author{Pochung Chen}
\affiliation{Department of Physics and Frontier Research Center on Fundamental and
Applied Sciences of Matters, \\
National Tsing Hua University, Hsinchu 30013, Taiwan}
\author{Ying-Jer Kao}
\affiliation{Department of Physics and Center for Advanced Study of Theoretical Science,\\
National Taiwan University, No. 1, Sec. 4, Roosevelt Rd., Taipei 10607,
Taiwan}
\author{Tao Xiang }
\affiliation{Institute of Physics, Chinese Academy of Sciences, P.O. Box 603, Beijing
100190, China}
\maketitle





\section{S1. Definition of quantum transfer matrix}

\begin{figure}[tb]
\includegraphics[width=8cm,clip]{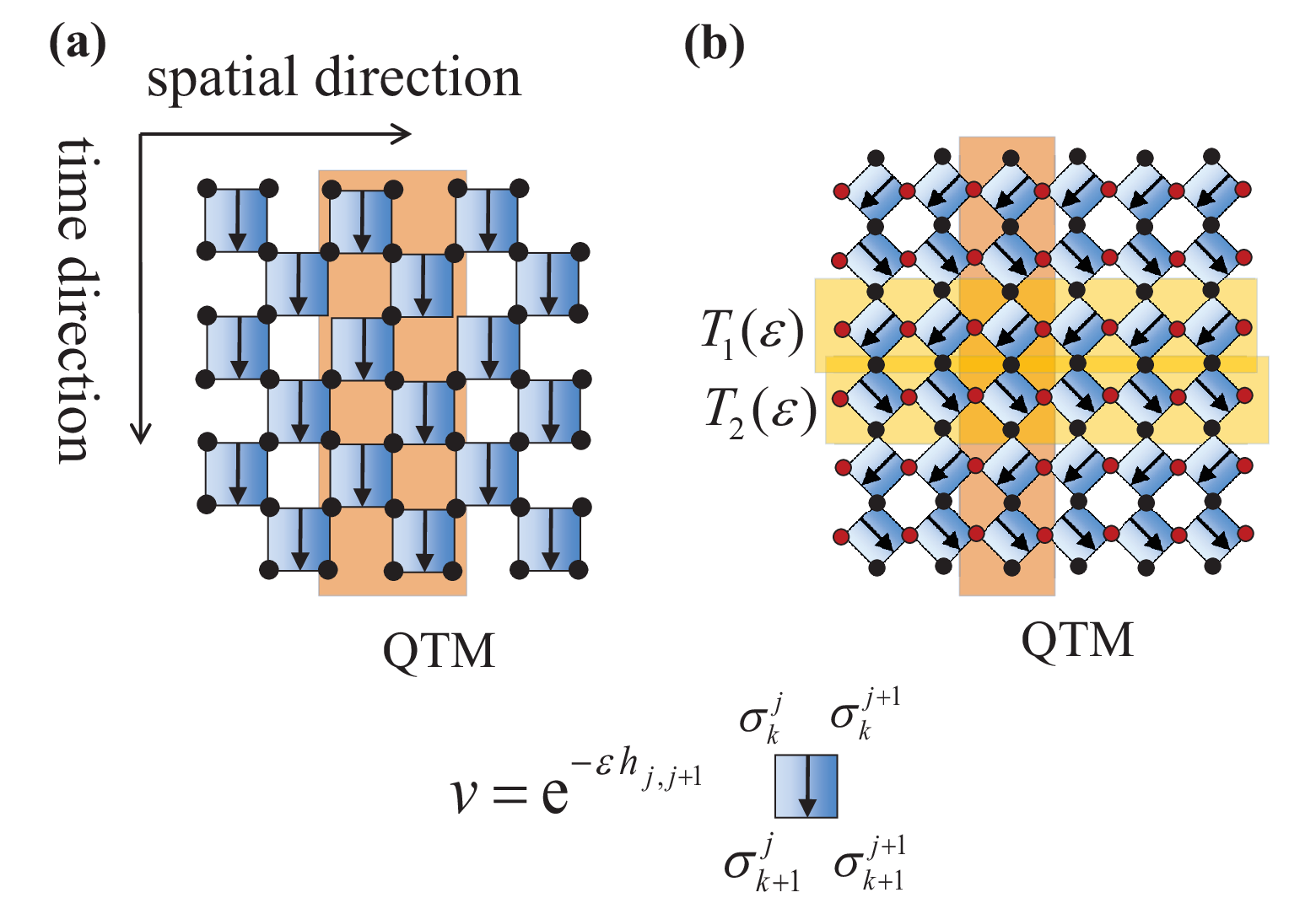} 
\caption{(Color online)
(a) From standard Trotter-Suzuki decomposition of the operator e$^{-1/T{H}}$, 
one can define a two-column wide QTM. 
(b) The mapping proposed in Ref.~\cite{Sirker:2002kl} maps e$^{-1/T{H}}$ onto
a tensor network with alternating rows and additional sites in an auxiliary 
mathematical space where one can define a one-column wide QTM.} \label{fig0}
\end{figure}

It is well-known that the standard quantum-to-classical mapping is the
Trotter-Suzuki decomposition which decomposes the thermal statistical
operator e$^{-1/T{H}}$ of a one-dimensional (1D) quantum system with
Hamiltonian ${H}=\sum_{j=1}^{L}{h}_{j,j+1}$ into a 2D tensor network showing
a checkerboard structure. Here $T$ indicates the temperature. As shown in
Fig. \ref{fig0}(a), this tensor network can be expressed as a product of
local transfer matrices $\nu \equiv $ e$^{-\varepsilon h_{j,j+1}}$ with
matrix element%
\begin{equation}
\nu _{k,k+1}^{j,j+1}=\left\langle \sigma _{k+1}^{j}\sigma
_{k+1}^{j+1}\right\vert \text{e}^{-\varepsilon h_{j,j+1}}\left\vert \sigma
_{k}^{j}\sigma _{k}^{j+1}\right\rangle ,
\end{equation}%
where $\varepsilon =1/MT$ is the imaginary-time step and $M$ the Trotter
number in the imaginary-time direction. The subscripts $j$ and $k$ represent
the physical site (represented graphically by full black circles)
coordinates in the spatial and imaginary-time directions, respectively. This
mapping has several disadvantages. The obvious one, for example, is that the
quantum transfer matrix (QTM) is two columns wide, leading to more memory
space and computational complexity. A more detailed discussions of its
disadvantages can be found in Ref. \cite{Sirker:2002kl}. In this work, we
employ an alternative quantum-to-classical mapping proposed in Ref. \cite%
{Sirker:2002kl} which decomposes the operator e$^{-1/T{H}}$ into e$^{-1/T{H}%
}\approx \underset{M\rightarrow \infty }{\lim }\left\{ \emph{T}%
_{1}(\varepsilon )\emph{T}_{2}(\varepsilon )\right\} ^{M/2}$ where $%
T_{1,2}(\varepsilon )=T_{R,L}$e$^{-\varepsilon H}$ and $T_{R}=$e$^{iP}$ ($%
T_{L}=$e$^{-iP}$), with $P$ being the momentum operator, denotes the right
(left) shift operator. The resulting 2D tensor network has alternating rows
and additional virtual (auxiliary) sites (represented graphically by full
red circles) as shown in Fig.~\ref{fig0}(b), where $T_{1}(\varepsilon )$ ($%
T_{2}(\varepsilon )$) is a row-to-row transfer matrix which is composed of a
45$^{\circ }$ clockwise (counter-clockwise) rotation of the local matrix $%
\nu $ and a one column wide QTM as indicated by a shaded rectangle can be
defined. Instead of acting on a chain of physical sites in the time
direction as in the Trotter-Suziki decomposition, this QTM acts on a chain
of virtual sites in the time direction. Such a mapping arises in the context
of exactly solvable model at finite temeprature \cite{Klumper:2004zr} and
Sirker \cite{Sirker:2002ys} has proved that this mapping has a second-order
error correction term and can be generally applied to 1D systems with
nearest neighbor interactions.

\begin{figure}[tb]
\includegraphics[width=8cm,clip]{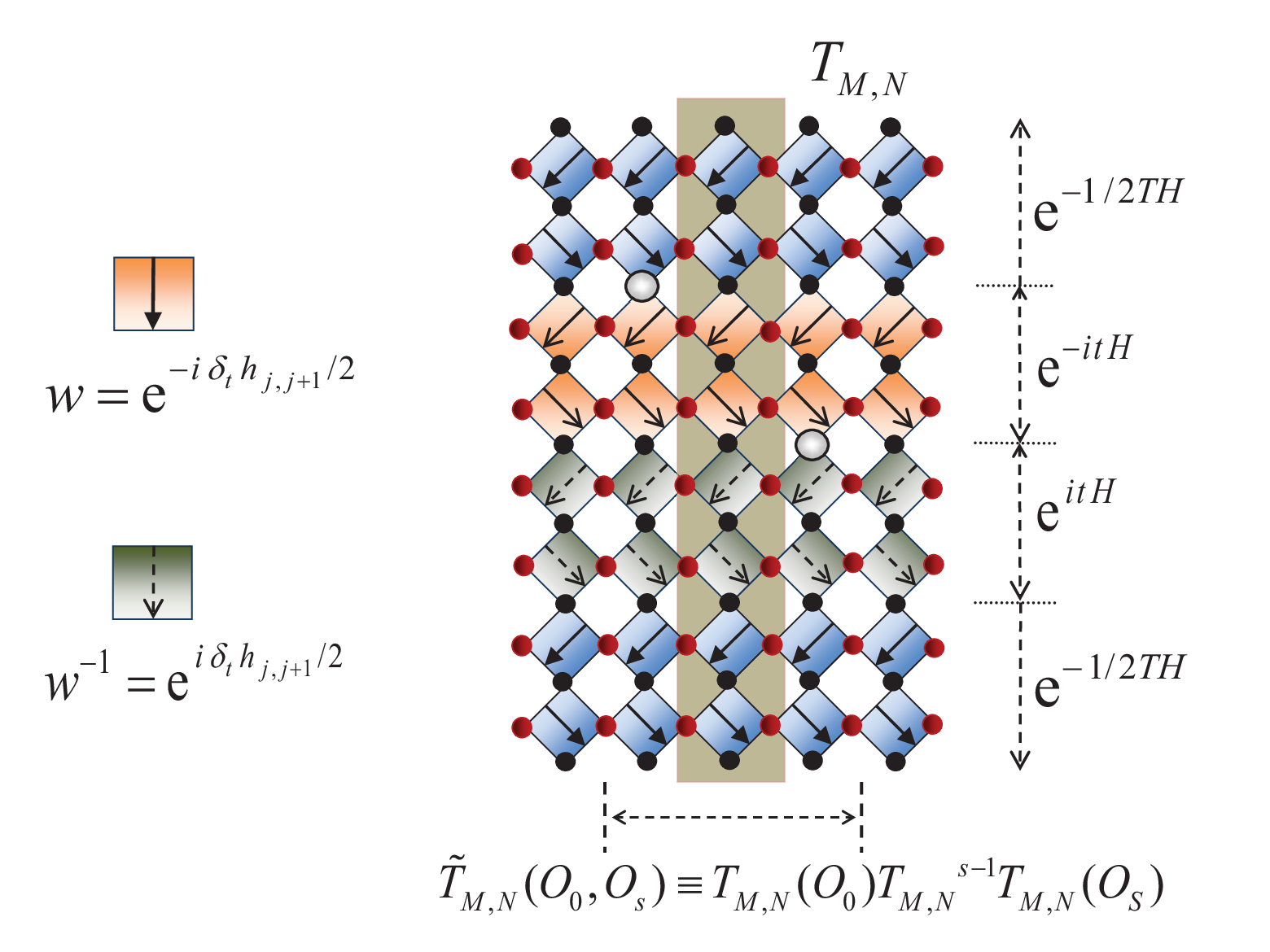} 
\caption{(Color online)
The alternative quantum-to-classical mapping decomposes the operator 
e$^{itH}$e$^{-itH}$e$^{-1/T{H}}$ of a 1D quantum system into a 2D tensor network where 
a QTM $T_{M,N}$ can be defined. The two-point dynamic correlation function 
involves a modified transfer matrix $\widetilde{T}_{M,N}(O_{0},O_{s})$ where 
the two observables at different site and different time are represented 
graphically by two empty circles.} \label{fig1}
\end{figure}

When considering the time-dependent correlation function as in Eq. (1) in
the main article, a similar decomposition can be applied to the
time-evolution operators e$^{-itH}$ and e$^{itH}$, with $\nu $ being
replaced by complex local matrices $w\equiv $ e$^{-i\delta _{t}h_{j,j+1}}$
and $w^{-1}\equiv $ e$^{i\delta _{t}h_{j,j+1}}$ respectively, and $%
\varepsilon $\ being replaced by the real-time step $\delta _{t}=t/N$ where $%
N$ denotes the Trotter number in the real-time direction. These additional
decompositions results in a 2D tensor network as shown in Fig. \ref{fig1}
where a new QTM $T_{M,N}$ can be defined. Due to the trace operation in Eq.
(1) in the main article, we note that this tensor network has a periodic
boundary condition at the top and bottom physical sites and an additional
transfer matrix $\widetilde{T}_{M,N}(O_{0},O_{s})\equiv
T_{M,N}(O_{0})T_{M,N}^{\text{ }s-1}T_{M,N}(O_{s})$ as shown in Fig. \ref%
{fig1} is involved, where $T_{M,N}(O_{j})$ is a modified transfer matrix
containing operator $O_{j}$ at site $j=0$ and $s$.

In the thermodynamic limit, the calculation of the dynamic correlation
function can be recast into Eq. (3) in the main article. As a result, two
main issues concerning the transfer-matrix renormalization group (TMRG)
calculation of the long-time dynamics for this correlator arise: the
identification of the entanglement structure of the maximal eigenvectors for
the $T_{M,N}$ and the efficient evaluation of the additional transfer matrix 
$\widetilde{T}_{M,N}(O_{0},O_{s})$ especially when the distance $s$ is
large. In the following, we shall discuss in details how to tackle these two
issues respectively.

\section{S2. Entanglement structure of the maximal eigenvector}

Let us first consider the case $T=\infty $ (e$^{-1/T{H}}=1$). Through the
quantum-to-classical mapping, the operator e$^{itH}$e$^{-itH}$ can be
decomposed into a 2D tensor network where a $M$-independent QTM $T_{N}$\ can
be defined, which is shown in Fig. \ref{fig2}. The discovery of the
entanglement structure of the maximal eigenvectors of $T_{N}$ provides us
not only the understanding that how the entanglement builds up in
conventional TMRG calculation but also a picture that how the entanglement
growth can be avoided by taking a different bi-partitioning scheme for the
QTM $T_{N}$. In this section,we will prove three important properties of $%
T_{N}$, from which one can identify the entanglement structure of the
maximal eigenvectors of $T_{N}$ as described in the main article. Assume the
dimension of the local Hilbert space is $d$. Let $\tau _{n}^{\ast }$ denote
a pair of virtual states $(\tau _{n},\overline{\tau }_{n})$ at the same
forward and backward real-time $n$, $n=1,..,N$, $|\tau _{n}^{\ast }\rangle
=\sum_{\tau _{n}\overline{\tau }_{n}}\delta _{\tau _{n},\overline{\tau }%
_{n}}|\tau _{n},\overline{\tau }_{n}\rangle $ a maximally entangled state of 
$\tau _{n}$ and $\overline{\tau }_{n}$, and $\langle \psi _{n}^{l}|$ and $%
|\psi _{n}^{r}\rangle $ the left and right maximal eigenvectors of $T_{n}$
respectively. These properties are based on a key observation as indicated
in Fig. \ref{fig2}. For an arbitrary $L$, contracting $T_{N}^{L}$ with $%
|\tau _{N}^{\ast }\rangle $ from the right, the two matrices $w$ and $w^{-1}$
enclosed in the shadow circle contract to an identity. Thus, two physical
sites ($\sigma _{N-1}$ and $\overline{\sigma }_{N-1}$) contract and two
virtual sites ($\tau _{N}$ and $\overline{\tau }_{N}$) at the left hand side
contract such that the left-adjacent matrices $w$ and $w^{-1}$ connects and
become an identity again. This process can continue from right to left
through the QTM, and finally we obtain a separated maximally entangled pair
state $|\tau _{N}^{\ast }\rangle $\ on the left. Two rows of local matrices,
enclosed by the dashed rectangle in Fig.~\ref{fig2}, contract to identities
and a QTM $T_{N-1}^{L}$ with one fewer Trotter number is formed. This can be
expressed as $T_{N}^{L}|\tau _{N}^{\ast }\rangle =|\tau _{N}^{\ast }\rangle
T_{N-1}^{L}.$

\begin{enumerate}
\item The maximal eigenvalue $\Lambda _{0}$\ of $T_{N}$ is $d$ irrespective
of the Trotter number $N$, i.e., 
\begin{equation}
\Lambda _{0}=d.  \label{eq:1}
\end{equation}

\begin{figure}[tb]
\includegraphics[width=8cm,clip]{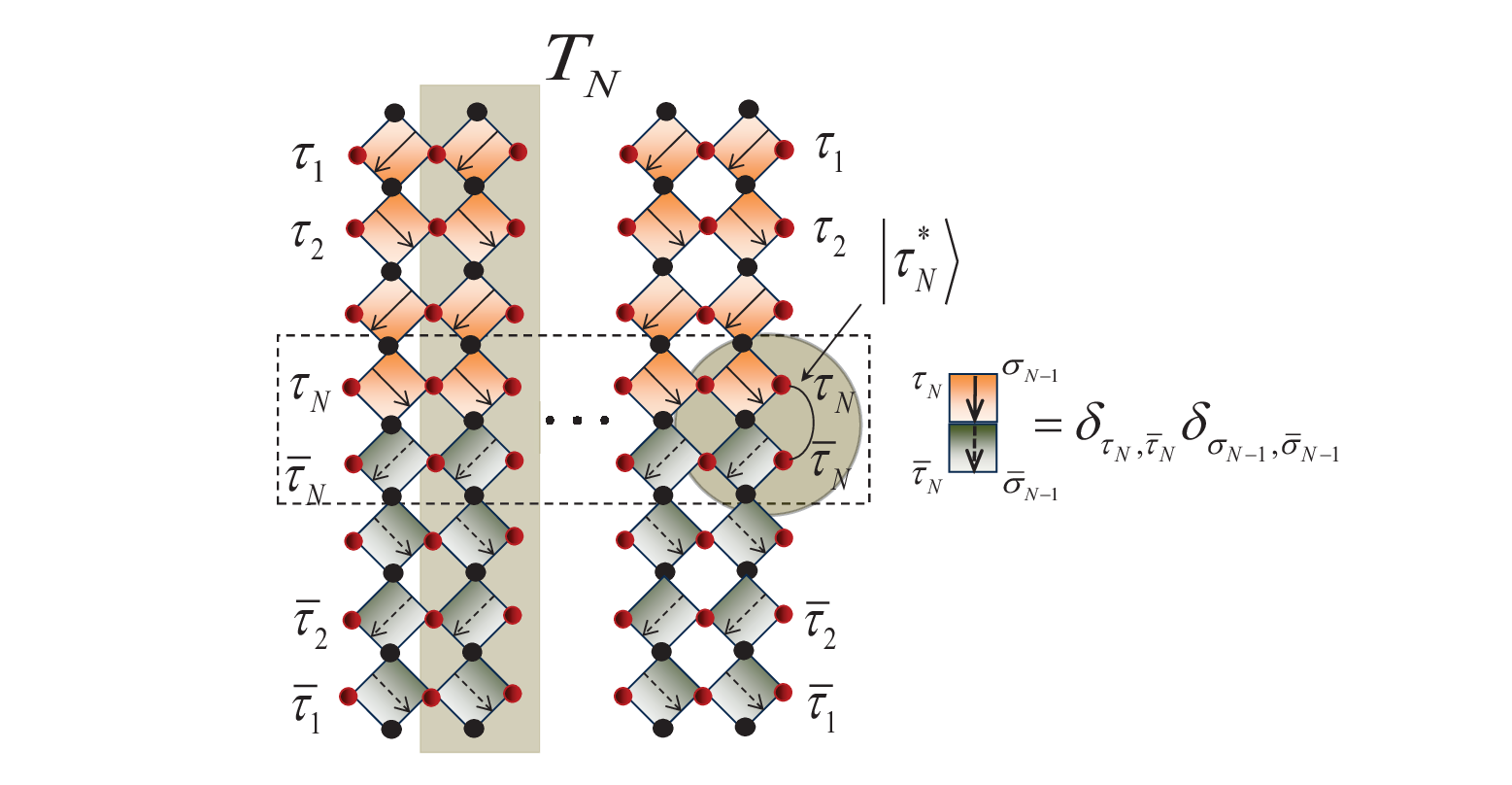} 
\caption{(Color online)
By contracting $T_{N}^{L}$ with $|\tau _{N}^{\ast }\rangle $ from the right, 
we have a tensor product of $|\tau _{N}^{\ast }\rangle $ on the left and 
$T_{N-1}^{L}$, which can be expressed as 
$T_{N}^{L}|\tau _{N}^{\ast }\rangle =|\tau _{N}^{\ast }\rangle T_{N-1}^{L}.$
} \label{fig2}
\end{figure}

\textit{proof: }It is straightforward to show that, as $L\rightarrow \infty $%
, $\Lambda _{0}^{L}=$Tr$(T_{N}^{L})=$Tr$($e$^{itH}$e$^{-itH})=d^{L}$.
Consequently, $\Lambda _{0}=d$ irrespective of the Trotter number $N$.
Furthermore, in the thermodynamic limit, one has 
\begin{equation}
\underset{L\rightarrow \infty }{\lim }T_{n}^{L}=d^{L}|\psi _{n}^{r}\rangle
\langle \psi _{n}^{l}|,
\end{equation}%
for arbitrary $n$. Here the normalization condition $\langle \psi
_{n}^{l}|\psi _{n}^{r}\rangle =1$ is used.

\item For all even $n$, the right eigenvector $|\psi _{n}^{r}\rangle $ is
related to the eigenvector $|\psi _{n-1}^{r}\rangle $\ by%
\begin{equation}
|\psi _{n}^{r}\rangle =|\tau _{n}^{\ast }\rangle |\psi _{n-1}^{r}\rangle .
\label{eq:2}
\end{equation}

\textit{proof: }The key observation in Fig. \ref{fig2} is a recurrence
relation. It satisfies the relation%
\begin{equation}
T_{n}^{L}|\tau _{n}^{\ast }\rangle =|\tau _{n}^{\ast }\rangle T_{n-1}^{L},
\end{equation}%
for all even Trotter number $n$ and arbitrary $L$. Therefore, one has%
\begin{eqnarray}
&&T_{n}^{L}|\tau _{n}^{\ast }\rangle |\psi _{n-1}^{r}\rangle  \notag \\
&=&|\tau _{n}^{\ast }\rangle T_{n-1}^{L}|\psi _{n-1}^{r}\rangle =d^{L}|\tau
_{n}^{\ast }\rangle |\psi _{n-1}^{r}\rangle .
\end{eqnarray}%
Accordingly, $|\tau _{n}^{\ast }\rangle |\psi _{n-1}^{r}\rangle $ is a right
eigenvector of $T_{n}$\ with eigenvalue $d$. Since $d$ is always the maximal
eigenvalue of $T_{n}$, one obtains the property (2) of the QTM $T_{N}$.

\item For all even $n$, by contracting the left eigenvector $\langle \psi
_{n}^{l}|$ with $|\tau _{n}^{\ast }\rangle $, we have%
\begin{equation}
\langle \psi _{n}^{l}|\tau _{n}^{\ast }\rangle =\langle \psi
_{n-1}^{l}|=\langle \psi _{n-2}^{l}|\langle \tau _{n-1}^{\ast }|,
\label{eq:3}
\end{equation}%
and the eigenvector $|\psi _{n-1}^{r}\rangle $\ has a similar property.

\textit{proof: }As $L\rightarrow \infty $, from $T_{n}^{L}|\tau _{n}^{\ast
}\rangle =|\tau _{n}^{\ast }\rangle T_{n-1}^{L}$ and $T_{n-1}^{L}=d^{L}|\psi
_{n-1}^{r}\rangle \langle \psi _{n-1}^{l}|$, one obtains 
\begin{equation}
T_{n}^{L}|\tau _{n}^{\ast }\rangle =d^{L}|\tau _{n}^{\ast }\rangle |\psi
_{n-1}^{r}\rangle \langle \psi _{n-1}^{l}|.  \label{eq:4}
\end{equation}%
Similarly, from $T_{n}^{L}|\tau _{n}^{\ast }\rangle =d^{L}|\psi
_{n}^{r}\rangle \langle \psi _{n}^{l}|\tau _{n}^{\ast }\rangle $ \ and Eq. (%
\ref{eq:2}), one obtains 
\begin{equation}
T_{n}^{L}|\tau _{n}^{\ast }\rangle =d^{L}|\tau _{n}^{\ast }\rangle |\psi
_{n-1}^{r}\rangle \langle \psi _{n}^{l}|\tau _{n}^{\ast }\rangle .
\label{eq:5}
\end{equation}%
By comparing Eq. (\ref{eq:4}) with Eq. (\ref{eq:5}), we conclude that%
\begin{equation}
\langle \psi _{n}^{l}|\tau _{n}^{\ast }\rangle =\langle \psi _{n-1}^{l}|.
\end{equation}%
It is easy to see that, however, $T_{n-1}^{L}$ has the same structure as $%
T_{n}^{L}$, with the roles of left and right eigenvectors interchange with
each other. Consequently, one has%
\begin{equation}
\langle \psi _{n-1}^{l}|=\langle \psi _{n-2}^{l}|\langle \tau _{n-1}^{\ast }|
\end{equation}%
according to Eq. (\ref{eq:2}). Similarly, because $T_{n-1}^{L}$ has the same
structure as $T_{n}^{L}$, all the argument applying to $\langle \psi
_{n}^{l}| $ also apply to $|\psi _{n-1}^{r}\rangle .$This is exactly the
property (3) of the QTM $T_{N}$.
\end{enumerate}

\begin{figure}[tb]
\includegraphics[width=8cm,clip]{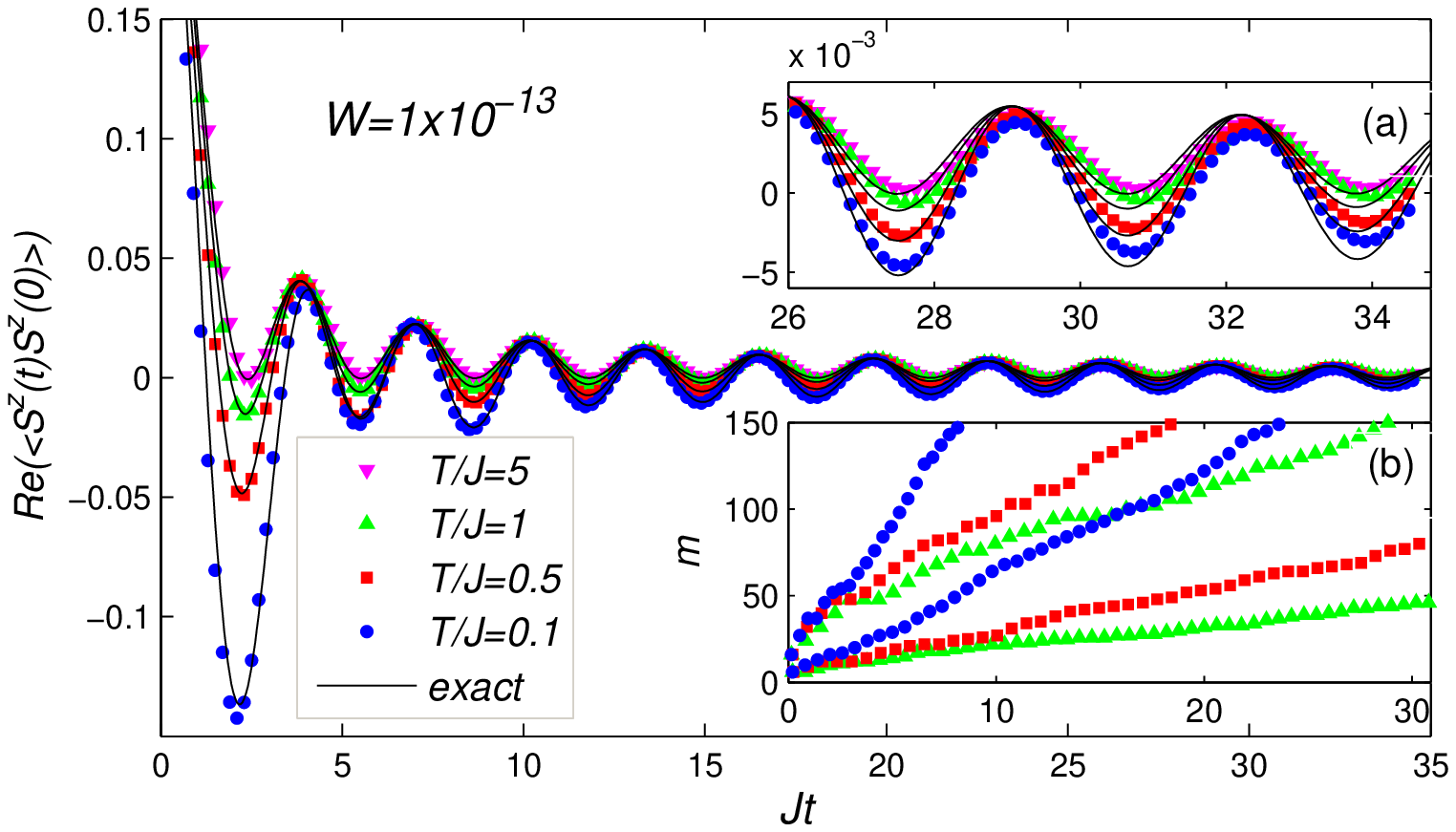} 
\caption{(Color online)
Autocorrelation Re($\langle S^{z}(t)S^{z}(0)\rangle$) of XXZ chain at various temperatures.
Inset (a): Blow-up of the Re($\langle S^{z}(t)S^{z}(0)\rangle$) near the large time-scale region.
Inset (b): The number of keeping states under fixed discarded
weight $W=10^{-13}$ increases linearly.} \label{figS1}
\end{figure}

From these properties, we argue that the maximal eigenvectors of $T_{N}$
have an entanglement structure as depicted in Fig. 2(b) in the main article.

As taking into account the finite-temperature effect, the QTM $T_{M,N}$ is
obtained by adding two blocks of thermal operator at the top and bottom of $%
T_{N}$ as shown in Fig. \ref{fig1}. Similar to the derivation of Eq. (\ref%
{eq:1}), it is easy to obtain%
\begin{equation}
\Lambda _{0}=\rho _{M},
\end{equation}%
where $\rho _{M}$ denotes the maximal eigenvalue of QTM $T_{M}$ involving
only imaginary-time. Meanwhile, since the key observation in Fig. \ref{fig2}
still hold for the finite-temperature case, .i.e., $T_{M,N}^{L}|\tau _{N}^{\ast
}\rangle =|\tau _{N}^{\ast }\rangle T_{M,N-1}^{L}.$ Following the above
derivation, for the maximal eigenvectors at finite temperaure, it can also
be obtained 
\begin{equation}
\langle \psi _{M,n}^{l}|\tau _{n}^{\ast }\rangle =\langle \psi
_{M,n-1}^{l}|=\langle \psi _{M,n-2}^{l}|\langle \tau _{n-1}^{\ast }|
\end{equation}%
for an even real-time Trotter number $n$, and the eigenvector $|\psi
_{M,n}^{r}\rangle $ has a similar property for odd $n$. Here $\langle \psi
_{M,n}^{l}|$ and $|\psi _{M,n}^{r}\rangle $ denote the left and right
maximal eigenvectors of $T_{M,n}$ with $M$ fixed and $n=1,...,N$. Thus, we
assert that the eigenvectors of $T_{M,N}$ have a slower increasing
entanglement between the block involving imaginary-time and the block
involving real-time. Furthermore, the eigenvectors of $T_{M,N}$ have a
similar entanglement structure for the block involving real-time as the
structure of the eigenvectors of $T_{N}$ as shown in Fig. 2(c) in the main
article.


\section{S3. Real-time dynamics at Finite and Zero Temperature}

Here we show our results for the spin-1/2 XXZ chain for $\Delta =0$ at
various finite temperatures. In Fig.~\ref{figS1} we plot the longitudinal
spin autocorrelation function Re($\langle S^{z}(t)S^{z}(0)\rangle $) as a
function of time at temperatures $T/J=5,1,0.5,$ and $0.1$ respectively. In
the inset Fig.~\ref{figS1}(a) we zoom in the large time regime. It is clear
that the results are still very accurate at large time. In the inset Fig.~%
\ref{figS1}(b) we plot the time-dependent number of keeping states $m$,
under fixed discarded weight $W=1\times 10^{-13}$. It shows a linear,
instead of exponential, growth with time. For each temperature, there are
two time-dependent $m$-lines corresponding to the cut through the dark and
light shadows respectively. Furthermore, we find that the number $m$
increases faster with time when the temperature decreases. These results are
consistent with our prediction of the entanglement structure for the maximal
eigenvectors at finite temperatures. As shown in the figure, we successfully
reach time scale $Jt=35 $ or more.

\begin{figure}[tb]
\includegraphics[width=8cm,clip]{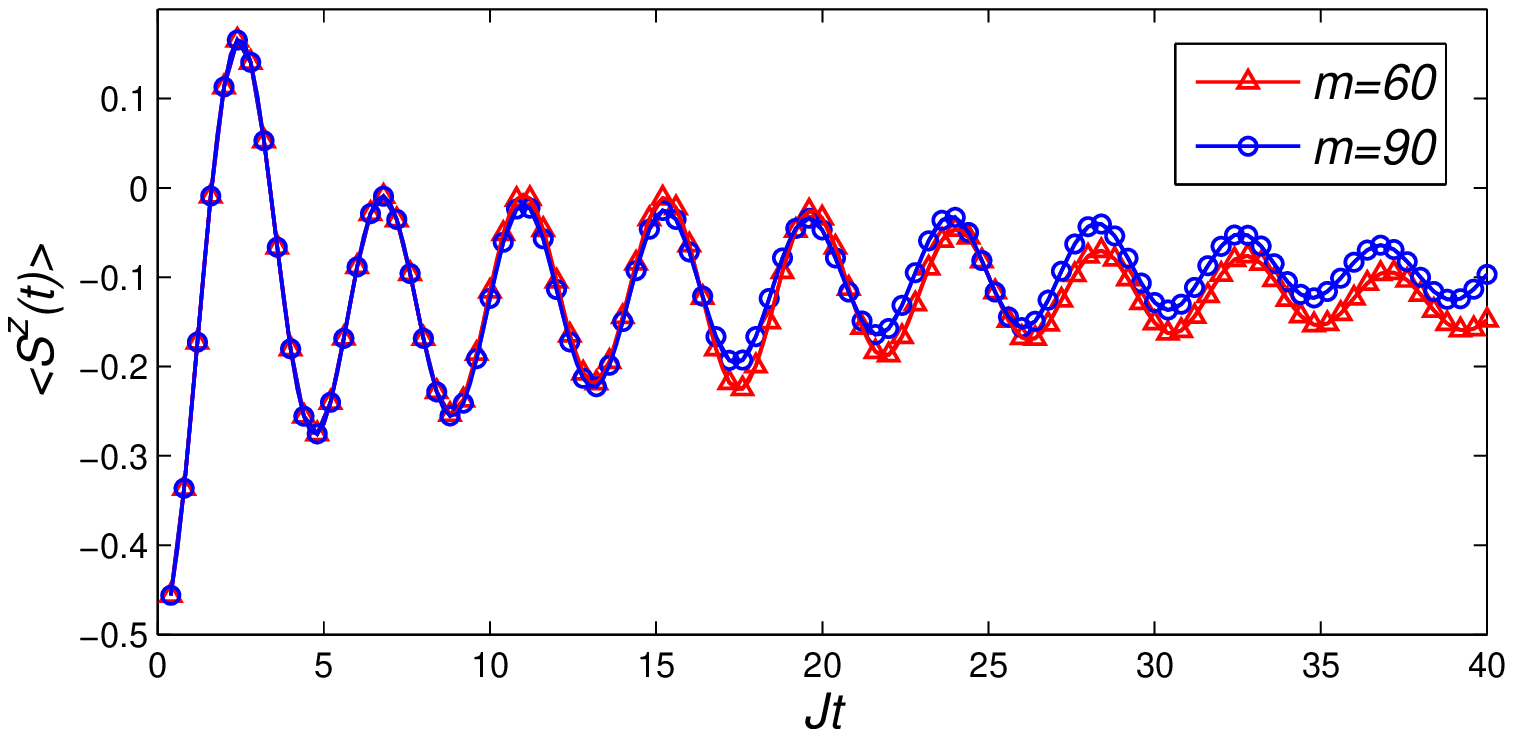} \vspace{0cm} 
\caption{(Color online)
Time-dependent magnetization of an XY chain at zero temperature with initial state
$|  0 \rangle =| \downarrow \rangle^{\otimes L}$.
Here, the time scale should be divided by a factor of $4$ to be comparable to
Fig. 12 in \cite{Banuls2}.
} \label{figS2}
\end{figure}

Our BTMRG scheme can also be applied to calculate the long-time dynamics of
quantum chains for pure states at zero-temperature. In this case, the upper
and lower boundary of the 2D tensor network (see Fig. 2(a) in the main
article) is restricted to fixed boundary conditions, i.e., contracting the
pure state at the upper and lower boundary of the network. As an example, we
calculate the time-evolution of magnetization per site $\langle
S^{z}(t)\rangle $ of an XY model at zero temperature with initial state $%
|0\rangle =|\downarrow \rangle ^{\otimes L}$. The Hamiltonian reads 
\begin{equation}
H=\sum_{j}J(S_{j}^{x}S_{j+1}^{x}-\frac{1}{2}S_{j}^{y}S_{j+1}^{y}).
\end{equation}%
The results with $m=60,90$ are shown in Fig. \ref{figS2}. Our results are
comparable with the results obtained by the folding algorithm (see Fig. 12
in \cite{Banuls2}, with caution that our time scale should be divided by a
factor of $4$ to be comparable to their definition of Hamiltonian).

\section{S4. Comparison between our BTMRG method and the modified tDMRG
method}

In this section, we provide details in dealing with the issue of efficient
evaluation of the large distance correlation function. Some comparisons
between our BTMRG method and the modified tDMRG method proposed in Ref.~\cite%
{Karrasch2012,*Barthel2012} are also given. 

In BTMRG framework, apart from the evaluation of the maximal eigenvalue $%
\Lambda _{0}$\ and the associated dual eigenvectors $|\psi ^{l}\rangle $ and 
$|\psi ^{r}\rangle $, the correlator as in Eq. (3) in the main article
involves the evaluation of the modified transfer matrix $\widetilde{T}%
_{M,N}(O_{0},O_{s})$ (see Fig. \ref{fig1}).

For autocorrelation function $\langle S^{z}(t)S^{z}(0)\rangle $ as shown in
Fig. \ref{figS1}, this additional transfer matrix is only one-column wide.
The correlator can be easily obtained by splitting $\widetilde{T}%
_{M,N}(S^{z},S^{z})$ into system and environment blocks, enlarging both
blocks, and projecting both blocks onto the dual reduced biorthonormal bases
describing the left $|\psi ^{l}\rangle $\ and right $|\psi ^{r}\rangle $
maximal eigenvectors, just as in the same way of the treatment for the QTM $%
T_{M,N}$. However, in certain cases, one has to calculate correlation
function for two operators far distance away. It is a challenge for BTMRG to
evaluate large-distance correlators since it involves the evaluation of a
wide transfer matrix $\widetilde{T}_{M,N}(O_{0},O_{s})$. Here, by
approximating $\widetilde{T}_{M,N}(O_{0},O_{s})$\ as a matrix-product
operator (MPO) and renormalizing the MPO matrices in a similar way as
renormalizing the QTM $T_{M,N}$, we succeed to calculate the correlator very
efficiently for large distance $s$. In particular, we show that our BTMRG is
equivalent to a novel tDMRG algorithm exactly in the thermodynamic limit.

\begin{figure}[tb]
\includegraphics[width=8cm,clip]{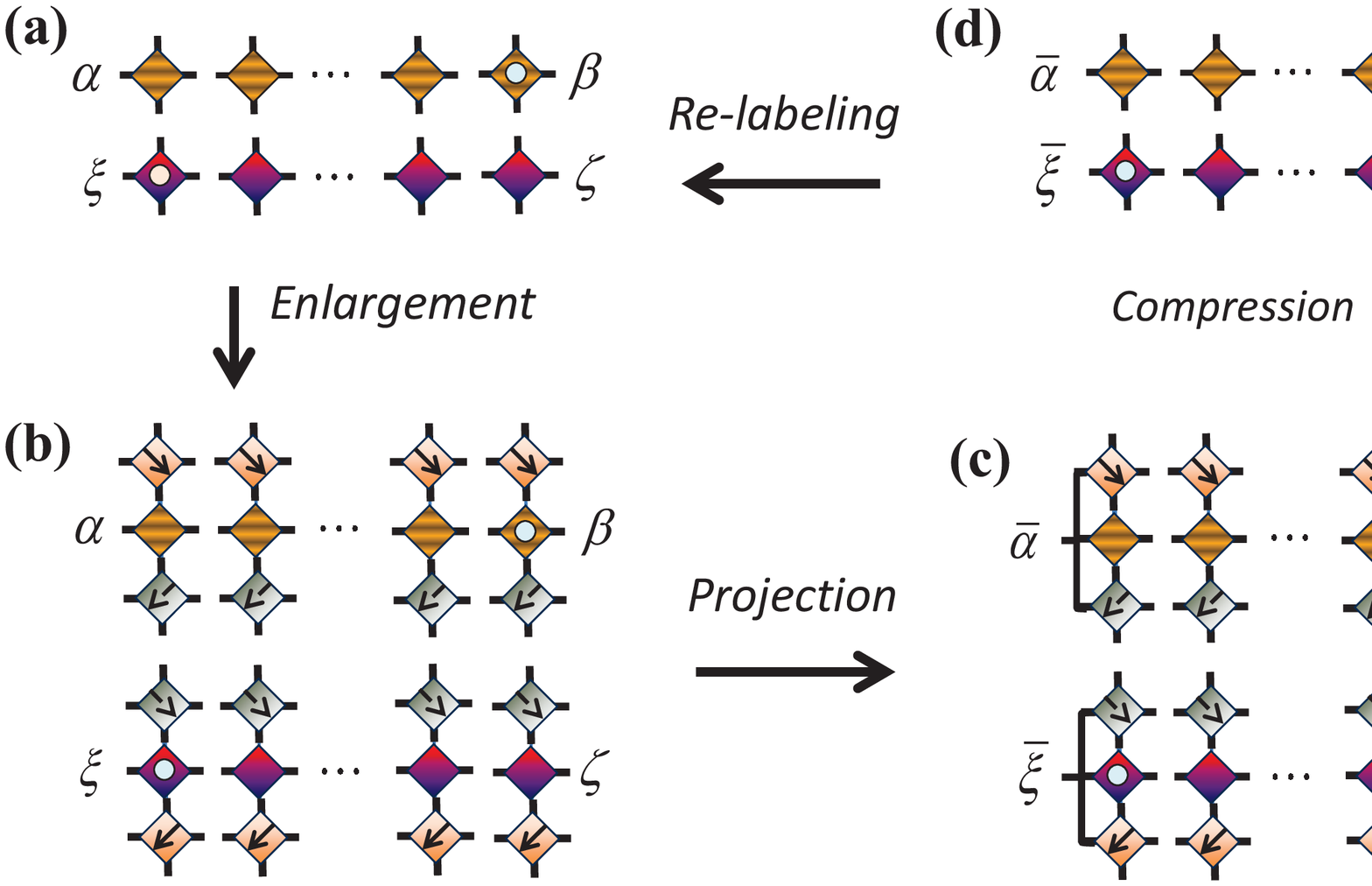} \vspace{0cm} 
\caption{(Color online)
Main steps of renormalization group for $\widetilde{T}_{M,N}(O_{0},O_{s})$.
(a) MPO representation of the system and environment blocks. The vertical indices
are the analogy of the physical quantum state and ancilla state in recent tDMRG algorithm.
$\{|\alpha \rangle \}$ and $\{|\beta \rangle \}$ ($\{|\xi \rangle \}$ and $\{|\zeta \rangle \}$)
represent the dual biorthonormal bases in describing the left $|\psi ^{l}\rangle $\
and right $|\psi ^{r}\rangle $ dominant eigenstates for system (environment) block.
(b) Enlarging MPO matrices by adding $w=$e$^{-i\delta _{t}h_{j,j+1}}$
and $w^{-1}=$ e$^{i\delta _{t}h_{j,j+1}}$ in between the system and environment blocks.
This process is equivalent to the main feature of tDMRG: evolving the quantum
state with ancilla state evolved in reverse time.
(c) Projecting the first and last enlarged MPO matrices from the left and right onto the new
biorthonormal basis states $\{|\overline{\alpha }\rangle \}$ and $\{|\overline{\beta }\rangle \}$ ($\{|\overline{\xi }\rangle \}$ and $\{|\overline{\zeta }\rangle \}$)\ for the enlarged system (environment) block obtained through BTMRG.
(d) Compressing the MPO by carrying out SVD and truncation on each bond of the MPO.
The correlation function is obtained by contracting the MPOs with $|\psi ^{l}\rangle $
and $|\psi^{r}\rangle $ at the left and right bonds of the MPOs.
} \label{figS3}
\end{figure}

In Fig. \ref{figS3} we sketch the main renormalization group (RG) steps for
the transfer matrix $\widetilde{T}_{M,N}(O_{0},O_{s})$ within BTMRG
framework. In each step, the transfer matrix $\widetilde{T}%
_{M,N}(O_{0},O_{s})$ is split into the system (the upper) and environment
(the lower) blocks according to our special bi-partitioning configuration.
Each block is expressed as a $(s+1)$-length MPO, where each column of the
block is represented as a local tensor (MPO matrix) with two vertical
indices labelling site basis states and two horizontal indices labelling
bond states between the MPO matrices.

We start from Fig. \ref{figS3}(a) where the left and right maximal
eigenvectors can be expressed in terms of the dual biorthonormal bases $%
\{|\alpha \rangle \}$ and $\{|\beta \rangle \}$ ($\{|\xi \rangle \}$ and $%
\{|\zeta \rangle \}$) for the system (environment) block. 
The next step (Fig. \ref{figS3}(b)) is to enlarge every MPO matrix by adding
local matrices $w=$ e$^{-i\delta _{t}h_{j,j+1}}$ and $w^{-1}=$ e$^{i\delta
_{t}h_{j,j+1}}$ in between the system and environment blocks. These two 45$%
^{\circ }$-rotations of $w$ and $w^{-1}$ alternate in the clockwise and
counter-clockwise manner for every two RG iterations.

In the mean time, on the other hand, the QTM $T_{M,N}$ undergoes the same
process and the new reduced dual biorthonormal bases $\{|\overline{\alpha }%
\rangle \}$ and $\{|\overline{\beta }\rangle \}$ ($\{|\overline{\xi }\rangle
\}$ and $\{|\overline{\zeta }\rangle \}$)\ for the enlarged system
(environment) block are obtained through BTMRG method \cite%
{YHuang2011a,*YHuang2011b,*YHuang2012}. Then, we first project the first and
last enlarged MPO matrices from the left and right onto the new basis states
respectively (Fig. \ref{figS3}(c)), and subsequently compress the MPO by
carrying out singular-value decomposition (SVD), keeping the $\chi $ most
relevant states, and truncating the other irrelevant states on each bond of
the MPO (Fig. \ref{figS3}(d)). As a consequence, the dynamic correlator is
evaluated by contracting the MPOs with the left $|\psi ^{l}\rangle $ and
right $|\psi ^{r}\rangle $ maximal eigenvectors at the left and right bonds
of the MPOs. This completes a cycle of the RG steps and the next cycle is
repeated by re-labelling the biorthonormal basis state: $\overline{\alpha }%
\rightarrow \alpha ,$ $\overline{\beta }\rightarrow \beta ,$ $\overline{\xi }%
\rightarrow \xi ,$ and $\overline{\zeta }\rightarrow \zeta $ until the
desired time scale is reached.

We particularly note that the vertical indices of the MPO in our BTMRG are
the analogy of the physical quantum state and ancilla state in recent tDMRG
algorithm \cite{Karrasch2012,*Barthel2012}. The process in step (b) is
completely equivalent to the main feature of tDMRG: evolving the quantum
state with ancilla state evolved in reverse time. Since tDMRG merely works
with finite system, our BTMRG algorithm can thus be regarded as a novel
tDMRG algorithm exactly in the thermodynamic limit. The most notable
advantages of our BTMRG over tDMRG are: (i) the infinite-length MPO in tDMRG
is reduced to a $(s+1)$-length MPO leading to dramatic saving of memory
space and computing time; (ii) the left (right) bond of the first (last) MPO
matrix of our BTMRG is obtained by projecting the MPO matrices from the left
(right) onto the new reduced basis states. Since the number $m$ of reduced
basis states is small (see Fig. \ref{figS1}), the bond dimension $\chi $ of
the MPO is thus smaller than that of tDMRG; and (iii) the dynamic correlator
is obtained by contracting the $(s+1)$-length MPO with the left and right
maximal eigenvectors at the left and right bonds respectively. In cases when
one has to calculate correlation functions for various distances, this
advantage is particularly significant because the left and right
eigenvectors only need to be calculated once while the MPS-evolution in
tDMRG must be performed from the beginning for each correlation function.


\bibliographystyle{apsrev4-1}
\bibliography{BTMRG}